\documentstyle[prb,aps,twocolumn]{revtex}

\begin{document}

\draft

\title{Statistics of fluctuations for two types of crossover:
from ballistic to diffusive regime and from orthogonal to unitary ensemble}

\author{Eugene Kogan and Moshe Kaveh }

\address{ Jack and Pearl Resnick Institute of Advanced Technology,\\
Department of Physics, Bar-Ilan University, Ramat-Gan 52900, Israel}

\date{\today}
\maketitle
\begin{abstract}
In our previous publication
[Kogan {\it et al}, Phys. Rev. {\bf 48}, 9404 (1993)]
we considered the issue  of  statistics  of radiation diffusively
propagating in a disordered medium. The consideration was
in the framework of diagrammatic
techniques and  a new representation for the intensity distribution
function in terms of connected diagrams only was proposed.
Here we use similar approach  to treat the issue of
 statistics in the regime of the crossover  between
 ballistic and  diffusive transport. We find that even small
contribution from coherent component decreases by one half the intensity
distribution function for
 small values of intensity and also produces oscillations
of the distribution function.
We also apply this method to study statistics of fluctuations of
wave functions of chaotic electrons
in a quantum dot in an arbitrary magnetic field,
 by calculating the single state
local density in the regime of the
crossover between the orthogonal and unitary ensemble.
\end{abstract}

\narrowtext

\section{Introduction}
When electromagnetic wave propagates in a random medium and undergoes
multiple scattering the density of energy in a given point (intensity) $%
I=<E^2>$ is a strongly fluctuating quantity. In a diffusive regime these
fluctuations are traditionally described by Rayleigh statistics \cite{good}.
The distribution function $P_R(I)$ is a simple negative exponential:
\begin{equation}
\label{1}P_R(I)={\frac 1{<I>}}\exp \left( -{\frac I{<I>}}\right) ,
\end{equation}
which corresponds to the equation for the moments:
\begin{equation}
\label{2}<I^n>=n!<I>^n.
\end{equation}
In Eqs. (\ref{1}) and (\ref{2}) the averaging is with respect to all
macroscopically equivalent but microscopically different scatterers
configurations. By diffusive we mean the regime when it is possible to
discard the ''coherent'' component of the field, that is

\begin{equation}
\label{xx}<E>=0.
\end{equation}
This assumption is true for the case $r/\ell \gg 1$, where $r$ is a distance
from the source to the observation point, and $\ell $ is a mean free path.
The aim of the present publication is
to get the distribution function for arbitrary relation between $r$ and $%
\ell $, when generally speaking the ''coherent'' component of the field
should also be taken into account.

Let us remind how the issue of intensity statistics is treated in the
framework of the traditional diagrammatic technique \cite{shap,spec1}. For the
sake of definitness consider the case when the point source
is situated at the origin; the intensity is measured in the point $r$. In
the diagrammatic representation $<I>$ is given by the diagrams with a pair
of wave propagators $G_{0r}^R$ and $G_{0r}^A$, summed with respect to all
possible interactions with the scatterers. The n-th moment $<I^n>$ is given
by the set of diagrams with n propagators $G_{0r}^R$ and n propagators $%
G_{0r}^A$. For us is important the following property of diagrams: if the
diagram consists from several disconnected parts then the contribution of
that diagram is equal to the product of contributions of disconnected parts,
so all the moments can be expressed at least in principle through the
contributions of connected diagrams only. To get Rayleigh statistics one
should ignore all connected diagrams save those consisting from a pair of
propagators (advanced and retarded)\cite{shap,spec1}.

Previously  \cite{spec1} we have formulated a
perturbation theory which systematically takes all connected diagrams into
account.
It is connected diagrams with more than two propagators which give the
deviation
from the simple Rayleigh statistics for large values of intensity (for
the tail of the
distribution function). In this paper we would be interested not in the tail
but
in the "body" of the distribution function. (Because the
deviation appears for the values of intensity which are inversely proportional
to
$r$ this limitation looks quite natural in the regime considered). Hence we
would discard connected diagrams with more than two propagators.
On the other hand previously we  ignored the diagrams which contain isolated
(dressed) propagators.
The ground for this is that such diagrams
give contribution of the order of $\exp (-r/\ell )$,
and in the fully developed
diffusive regime one can ignore them. Here we should generally speaking
take them into account.

\section{Crossover between diffusive and ballistic regime}
For $<I^n>$ we consider only diagrams which
consists of isolated (dressed) propagators and pairs of propagators (advanced
and
retarded one). The sum of all diagrams can be written down in the following
way:
\begin{equation}
\label{3}<I^n>=\sum_{m=0}^nP(n,m)A^mB^{n-m},
\end{equation}
where $A=\left| <E>\right| ^2$, $B=<I>-|<E>|^2$, and $P(n,m)$ is the
coefficient of purely combinatorial origin:
\begin{equation}
\label{4}P(n,m)={\frac{(n!)^2}{(m!)^2(n-m)!}.}
\end{equation}

The distribution function $P(I)$ is connected with its moments by usual way:
\begin{equation}
\label{6}P(I)=\int_{-\infty }^\infty \exp (i\xi I)~\sum_{n=0}^\infty {\frac{%
(-i\xi )^n}{n!}}<I^n>~{\frac{d\xi }{2\pi }}.
\end{equation}
Substituting the Eq. (\ref{3}) into the Eq. (\ref{6}) and changing the order
of summation,which is possible due to absolute convergence of the series, we
get:

\begin{eqnarray}
\label{5a}
P(I)=\int_{-\infty }^\infty \exp (i\xi I)~\sum_{m=0}^\infty {\frac{%
\left( -i\xi A\right) ^m}{(m!)^2}} \nonumber\\
\times
\sum_{n=m}^\infty {\frac{n!}{{(n-m)!}}%
\left( -i\xi B\right) ^{n-m}}~{\frac{d\xi }{2\pi }}.
\end{eqnarray}
Taking into account that for $n\geq m$

\begin{equation}
\label{5b}{\frac{n!}{{(n-m)!}}X^{n-m}=\ \ \frac{d^m}{dX^m}{X}^n,}
\end{equation}
we finally obtain the distribution function in the form:
\begin{equation}
\label{9}P(I)=\int_{-\infty }^\infty {\frac{\exp \left( {i\xi I-{\frac{i\xi a
}{1+i\xi (1-a)}}}\right) }{1+i\xi (1-a)}}~~{\frac{d\xi }{2\pi }}
\end{equation}
(the intensity is measured in the units of $<I>$ and  parameter $%
a=A/(A+B)=|<E>|^2/<I>$,  is introduced). It is obvious that for $a=0$
(purely diffusive transport), the
Eq. (\ref{9}) coincides with  Eq. (\ref{1}), so this parameter gives the
degree of deviation from Rayleigh statistics. In the opposite limiting case $%
a=1$  Eq. (\ref{9}) gives
$P\left( I\right) =\delta \left( I-1\right) $.
This limiting case means that there are no fluctuations,
which can occur only when there is no scattering at all, in other
words in the ballistic regime.

For intermediate
value of $a$ the distribution function can easily be calculated
numerically.  This behavior is represented  on Fig. 1. We see that even for
 $a\ll 1$ (almost diffusive transport)  taking into account of the ''coherent''
component of the  field drastically changes the distribution function
 for small $I$; in
particular for $I=0$  we get: $P\cong 1/2$. Also interesting are the predicted
oscillations of the distribution function.

\section{Crossover between orthogonal and unitary ensemble}
Now we want to consider a second problem, which though physically different
from
the considered above is as we would see  very  similar mathematically. It's the
problem of statistics of fluctuations of wave functions of chaotic electrons
in a quantum dot in an arbitrary magnetic field. This problem
has been a subject of several studies \cite{wegner,altshuler,efetov1,efetov2}.
The distribution function $P_e\left( I\right) $ of a single state local
electron density  $I=|\psi \left( r\right) |^2$
was shown to be  for unitary ensemble (strong magnetic field)
 a simple negative exponent (like in Eq. (\ref{1})),
and for orthogonal ensemble (no magnetic field)
\begin{equation}
\label{ort}P_e(I)=\sqrt{\frac {<I>}{2 \pi I}}\exp \left( -{\frac I{2
<I>}}\right).
\end{equation}
In the paper \cite{efetov2} the distribution
function was calculated for the case of arbitrary magnetic field using the
supersymmetry technique. Let us see what the simple method  gives in
application
to this problem. Both limiting
cases can be easily obtained using the method of moments, if we
postulate for the density moment
\begin{equation}
<I^n>=<\underbrace{\psi \ldots \psi}_{n}
\,\,\underbrace{\psi ^{\ast}\ldots \psi ^{\ast}}_{n}> \nonumber
\end{equation}
the existence of Wick theorem. Then we obtain
\begin{equation}
\label{12u}<I^n>=n!<\psi \psi ^{\ast}>^n,
\end{equation}
where the multiplier $n!$ is simply the number of possible
pairings between amplitudes, and hence Rayleigh statistics
 For orthogonal ensemble the wavefunction is
real, that is there is no difference between $\psi$ and $\psi^{\ast} $ any
two propagators can be coupled and instead of  Eq.~(\ref{12u}) we get :
\begin{equation}
\label{12o}
<\underbrace{\psi \ldots \psi}_{2n}>=\left( 2n-1\right) !!<\psi \psi >^n,
\end{equation}
which gives us  Eq. (\ref{ort}).
(This statistics, by the way, is also known in the theory of
 propagation of classical waves in a disordered medium;
Eq. (\ref{ort}) gives distribution function for the case
of acoustic waves).
This gives us the idea how the general issue of electron density distribution
function should be addressed. We should take into account both the averages
$<\psi \psi ^{\ast}>$ and $<\psi \psi >$
(not supposing of course them to be equal).
 Then taking into account all possible ways
of coupling and using simple combinatorics  we get:
\begin{equation}
\label{13}<I^n>=
\sum_{m=0}^{\left[ n/2\right] }P(n,m)\left| <\psi \psi >\right|^m
<\psi \psi ^{\ast}>^{n-m},
\end{equation}
where $P(n,m)$ is the coefficient
of purely combinatorial origin:
\begin{equation}
\label{14}P(n,m)={\frac{(n!)^2}{(m!)^2(n-2m)!2^{2m}}.}
\end{equation}

Substituting Eq. (\ref{13}) into Eq. (\ref{6}) and  changing the order of
summation  we get:
\begin{eqnarray}
\label{5o}
P_e(I)=\int_{-\infty }^{\infty} \exp (i\xi I)~\sum_{m=0}^{\infty}
{\frac{\left( -i \xi \left| <\psi \psi >\right|\right) ^m}
{2^{2m}(m!)^2}}
\nonumber\\
\times
\sum_{n=2m}^\infty {\frac{n!}{{(n-2m)!}}%
\left( -i\xi <\psi \psi ^{\ast}>\right) ^{n-m}}~{\frac{d\xi }{2\pi }}.
\end{eqnarray}
Using   Eq.  (\ref{5b}) we obtain:
\begin{equation}
\label{19}P_e(I)={\int_{-\infty }^\infty \frac{\exp \left( {i\xi I}\right) }{
\sqrt{1+2i\xi -\xi ^2\;Y/(1+Y)}}}~~{\frac{d\xi }{2\pi }.}
\end{equation}
where the density is measured in the units of $<I>$
and $Y=<\psi \psi ^{\ast}>/\left| <\psi \psi >\right|-1$ is the
parameter which shows the character of the ensemble (for orthogonal ensemble
$Y=0$ and for unitary $Y=\infty $).
It is obvious that for $Y=\infty $ we get  Eq. (\ref{1}) and for $Y=0$ we get
Eq. (\ref{ort}).

For intermediate value of $Y$ the distribution function can easily be
calculated numerically. To compare our result with the
results of Ref. \onlinecite{efetov2},
on Fig. 2 we plotted the distribution $\varphi
\left( \tau \right) =2\tau P_e\left( \tau ^2\right) $ as function of $X =
\sqrt{%
Y}$ (looking at $I\rightarrow 0$  $Y\rightarrow 0$ asimptotics of Eq.
(\ref{19}) we
see that apart from numerical coefficient very close to 1 our parameter $X$
is equal to that introduced Ref. \onlinecite{efetov2}). Comparing our Fig. 2
with Fig. 1 of Ref. \onlinecite{efetov2} we see that they look very much
alike, though analytical expression for distribution function we have got
differs from that of  Ref. \onlinecite{efetov2}. \

\section*{Acknowledgements}
The authors acknowledge the financial support of the
Israeli Academy of Sciences.\

\begin{figure}
\caption{Distribution function for: (1) $a = .1$ (solid line),
(2) $a = .5$ (dashed line),
(3) $a = .9$ (dot-dashed line).}

\end{figure}

\begin{figure}
\caption{The distribution function $\varphi (\tau )$ of the local
 amplitude $\tau $ shown in the crossover regime for different values
of the parameter $X.$}
\end{figure} \

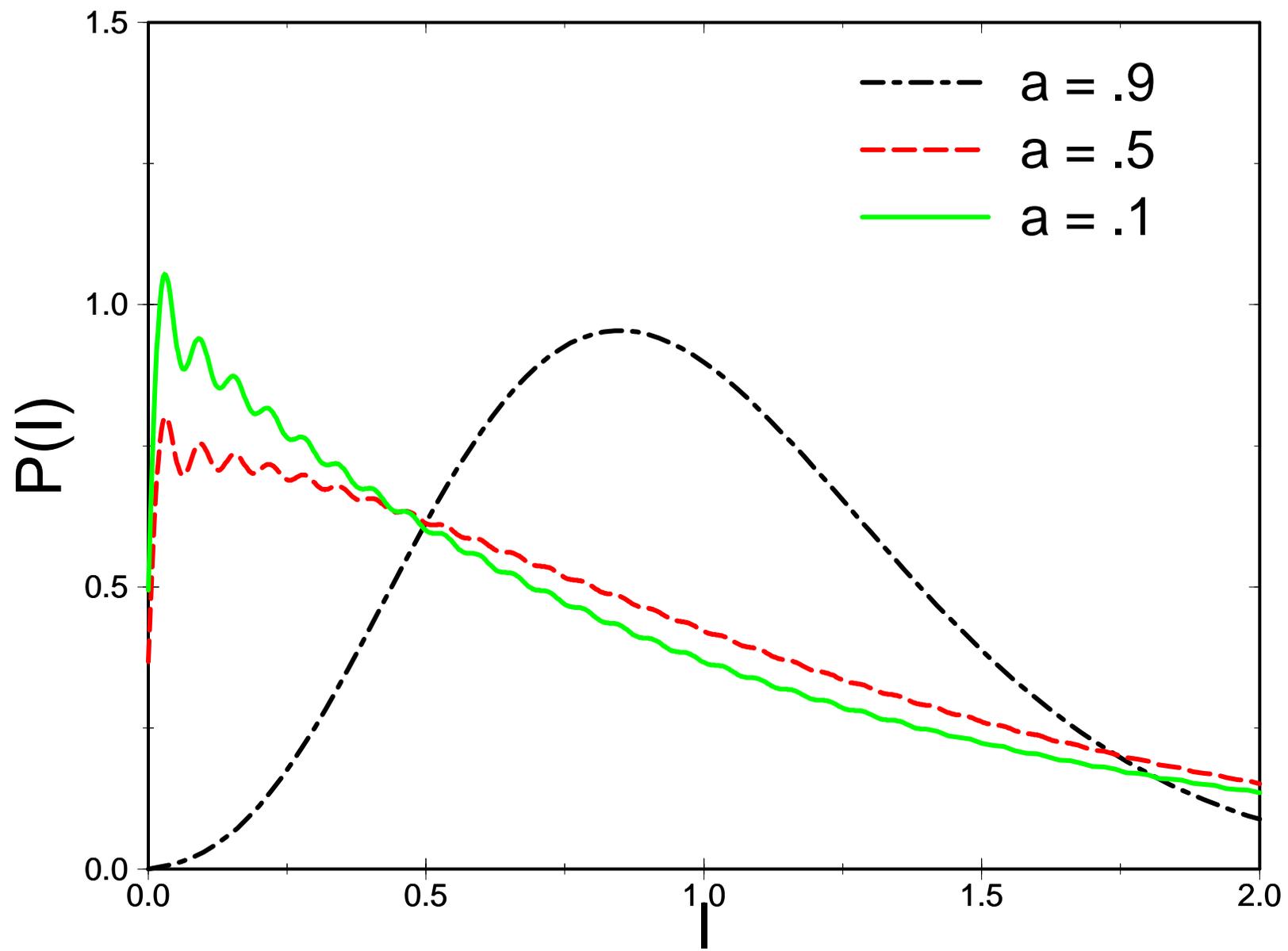
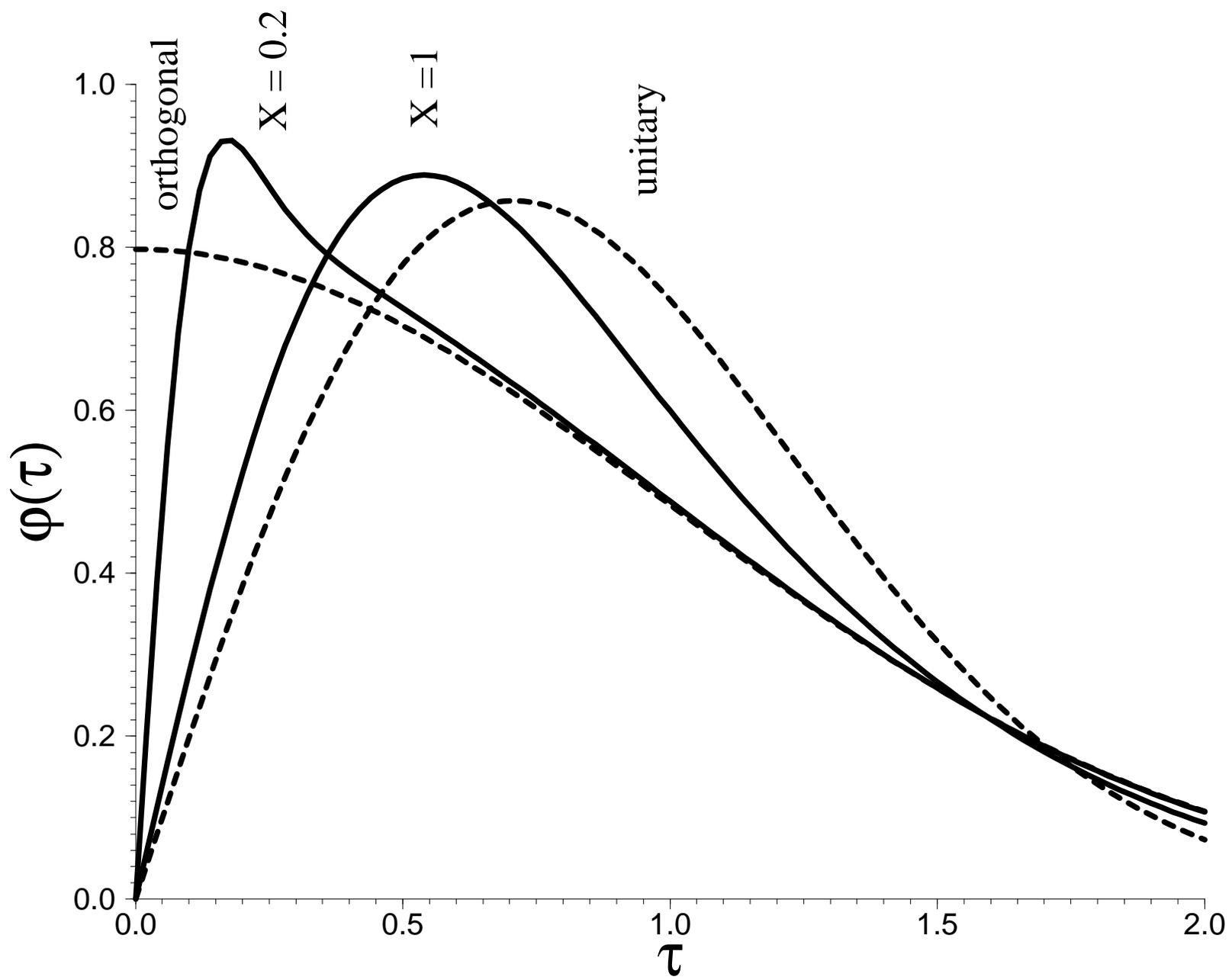
\end{document}